\documentclass{WS-PROCSm}
\usepackage{hyperref}

\begin{document}

\title{On the possible role of superheavy particles in the early
Universe\footnote{
\uppercase{T}his work is supported
by \uppercase{M}in. of \uppercase{E}ducation of \uppercase{R}ussia,
grant \uppercase{E}02-3.1-198.}}

\author{A. A. Grib}

\address{
A.Friedmann Laboratory for Theoretical Physics,     \\
Institute of Gravitation and Cosmology, PFUR.       \\
30/32 Griboedov kanal, St.Petersburg, 191023, Russia  \\
    Email:  andrei\_grib@mail.su    }

\author{Yu. V. Pavlov}

\address{
Institute of Mechanical Engineering, Russian Academy of Sciences, \\
 61 Bolshoy pr., V.O., St.Petersburg, 199178, Russia    \\
    Email:  yuri.pavlov@mail.ru   }


\maketitle

\abstracts{
    Different models of the role of creation of superheavy particles
in the early Friedmann Universe with their subsequent decay
on light particles are investigated.
    The observable numbers of baryon and entropy are predicted.
    The possible role of superheavy particles in creation of
cold dark matter is discussed.
}

\section{Introduction}

   It is known\cite{GMM,Grib,GD} that the number of particles with mass
of the order of the Grand Unification scale created by gravitation in
the early Universe described by the radiation dominated Friedmann metric
is of the Dirac-Eddington order, i.e. of the observable order for the
visible mass.
   Euristic considerations for particle creation in the
early Friedmann Universe leading to the prediction that the number of
created pairs of particles and antiparticles qualitatively is estimated
as the number of causally disconnected parts of space expanding to
the present size of horizon\cite{GD}
say that in spite of difficulties of exact calculations for the case
different from the radiation dominated regime the result must be of
the same order.
   On the other side it is clear that if superheavy particles after
their creation continued to be stable for large enough time they will
lead to the  collapse of the Universe governed by the radiation
dominated  metric in the short on the cosmological scale time for
closed  Friedmann space or lead to the unrealistic scale factor for
the open space.
   So the idea was proposed that these superheavy particles
must decay on quarks and leptons with $CP$-noninvariance leading to
the observable baryon charge of the Universe before the time when
the energy density of the created superheavy particles will become
equal to that creating the background metric.
    If superheavy particles have nonzero baryon charge then their
decay in analogy with decay of neutral $K$-mesons will go as decay of
some short living and long living components.
   Supposing that  the lifetime of long living components is of
the cosmological order but their number was diminished in comparison
with the number of the short living components due to their interaction
with the baryon charge created previously similar to the well known
regeneration mechanism for $K$-mesons one can speculate about their
existence today as cold dark matter.
   Rare events of their decays can be identified as experimental
observations of high energetic cosmic rays\cite{Takeda} with
the energy higher than the Greisen-Zatsepin-Kuzmin limit.\cite{GZK}
    Here we shall discuss different possibilities of the role of
superheavy particles with the mass of the Grand Unification scale in
the early Universe.

   1) It is natural to think that some inflation era took place before
the Friedmann stage.
   Some inflaton field which may be is manifesting
itself as the quintessence in the modern epoch after the quasi de~Sitter
stage led to the dust like or to the radiation dominated
Friedmann Universe.
    Usually it is supposed that the inflaton field does not interact
with ordinary particles and can be some manifestation of
the non Einsteinian gravity for example due to high order corrections.
   So even if it decayed on some light ``inflaton'' particles
the primordial inflaton field can form hot dark matter but not the
visible matter and entropy present in background radiation.
    Our idea is that inflaton field was the source of Friedmann metric
with some small inhomogeneities, but visible matter and the
entropy of the Universe were created not by the inflaton field
itself but by the gravitation of this inflaton field.
   That gravitation created pairs of superheavy particles.
   Short living components decayed in time of the Grand Unification scale
and led to the nonzero baryon charge observed today as visible matter.
   If long living components had the lifetime of the order of
the ``early recombination era'' then the  energy density of created
long living particles soon became equal to that of the background inflaton
field (hot dark matter).
   Then the decay of all long living components led to the observable
entropy of the Universe.
   Here it is supposed that the energy density of the inflaton  field
led to the observed cosmological scale factor, so it is evident that
the created entropy due to our mechanism will be of the observable order.

   2) The other possibility is to put the hypothesis discussed by us
earlier\cite{GrPv,GrPv2} that not all long living components decayed and
formed the entropy but some part of them survived up to modern time as
cold dark matter and superheavy particles  are  observed in
cosmic rays events.
   Then it is natural to suppose that the lifetime of the long living
component is of the cosmological order but the large part of them
regenerated into short living components  due to interaction with
the baryon charge in time shorter or equal to that of the
``early recombination era'' and entropy appeared due to this decay.

    Now let us give some numerical estimates.

\section{Model and Numerical Estimates}

     Total number of massive particles created in
Friedmann radiation dominated Universe
(scale factor $a(t)=a_0\, t^{1/2}$)      inside the horizon is as it
is known:\cite{GMM}
    \begin{equation}
N=n^{(s)}(t)\,a^3(t)=b^{(s)}\,M^{3/2}\,a_0^3 \ ,
\label{NbM}
\end{equation}
   where $b^{(0)} \approx 5.3 \cdot 10^{-4}$ for scalar
and  $b^{(1/2)} \approx 3.9 \cdot 10^{-3}$ for spinor particles
($ N \sim 10^{80} $ for $ M \sim 10^{14} $\,Gev, see Ref.~\refcite{GMM}).
    For the time ${t \gg M^{-1}} $ there is an era of going from the
radiation dominated model to the dust model of superheavy particles
    \begin{equation}
t_X\approx \left(\frac{3}{64 \pi \, b^{(s)}}\right)^2
\left(\frac{M_{Pl}}{M}\right)^4 \frac{1}{M}  \,.
\end{equation}
    If $M \sim 10^{14} $\,Gev,
$\ t_X \sim 10^{-15} $\,sec for scalar and
$\ t_X \sim 10^{-17} $\,sec for spinor particles.
   Let us call $t_X$ -- ``early recombination era''.

   Let us define $d $ --- the permitted part of long living
$X$-particles --- from the condition: on the moment of
recombination $t_{rec} $ in the observable Universe one has
$
d\,\varepsilon_X(t_{rec}) =\varepsilon_{crit}(t_{rec})  \,,
$
where $\varepsilon_{crit}$ is the critical density for the time $t_{rec}$.
    It leads to
\begin{equation}
d=\frac{3}{64 \pi \, b^{(s)}}\left(\frac{M_{Pl}}{M}\right)^2
\frac{1}{\sqrt{M\,t_{rec}}}\, .
\label{d}
\end{equation}
   For $M=10^{13} - 10^{14} $\,Gev one has
$d \approx 10^{-12} - 10^{-14} $ for scalar and
$d \approx 10^{-13} - 10^{-15} $ for spinor particles.
     So the life time of main part or all $X$-particles must be smaller
or equal than $t_X$.

    Now let us construct the model which can give: \
a) short living $X$-particles decay in time
   $\tau_q < t_X $ (more wishful is
   $\tau_q \sim t_C \approx 10^{-38} - 10^{-35} $\,sec,
i.e. Compton time for $X$-particles) \
b) long living particles decay with $\tau_l \approx t_X $.
   Baryon charge nonconservation with $CP$-nonconservation in full
analogy with the $K^0$-meson theory with nonconserved hypercharge and
$CP$-nonconservation leads to the effective Hamiltonian of the decaying
$X, \bar{X}$ - particles with nonhermitean matrix.

  For the matrix of the effective Hamiltonian
$ H=\{ H_{ij} \}, \ {i,j=1,2}$  let $H_{11}\! =\! H_{22}$
due to $CPT$-invariance.
    Denote
$\ \varepsilon=(\sqrt{\vphantom{ }H_{12}} - \sqrt{H_{21}}\,)\, / \,
(\sqrt{H_{12}} + \sqrt{H_{21}} \, )$.
    The eigenvalues $\lambda_{1,2} $ and eigenvectors
$|\Psi_{1,2}\rangle $  of matrix $H$ are
    \begin{equation}
\lambda_{1,2} = H_{11} \pm \frac{H_{12}+H_{21}}{2} \,
\frac{1-\varepsilon^2}{1+\varepsilon^2} \,,
\end{equation}
    \begin{equation}
|\Psi_{1,2}\rangle =\frac{1}{\sqrt{2\,(1+|\varepsilon |^2)}}\,
\left[ (1+\varepsilon) \,|1\rangle \pm \,(1- \varepsilon) \,
 |2\rangle \right].
\end{equation}
         In particular
\begin{equation}
H=     \left(
\begin{array}{cc}
E-\frac{i}{4}\left(\tau_q^{-1} +\tau_l^{-1}\right)
  &
\frac{1+\varepsilon}{1-\varepsilon}
\left[A-\frac{i}{4}\left(\tau_q^{-1} -\tau_l^{-1}\right)\right]
 \\  & \\
\frac{1-\varepsilon}{1+\varepsilon}
\left[A-\frac{i}{4}\left(\tau_q^{-1} -\tau_l^{-1}\right)\right]
 &
E-\frac{i}{4}\left(\tau_q^{-1} +\tau_l^{-1}\right) \\
\end{array}        \right) .
\label{HM}
\end{equation}

    Then the state $|\Psi_1 \rangle $ describes
short living particles $X_q$ with the life time
$ \ \tau_q \ $ and mass $E+A$.
    The state $\ |\Psi_2 \rangle $ is the state of long living particles
$X_l$ with life time $ \tau_l \ $  and mass $E-A$.
    Here $A$ is the arbitrary parameter $-E<A<E$  and it can be zero,
$E=M$.

    So for the scenario~1) it is sufficient to take
$\tau_l \approx t_X$.

    In scenario~2)
the small $ d \sim 10^{-15} - 10^{-12} $ part of long living
   $X$-particles with $\tau_l > t_U \approx 10^{18}$\,sec
   \  ($t_U $ is the age of the Universe)
is forming the dark matter.
   The decay of these superheavy particles in modern epoch
can give observed ultra high energy cosmic rays.
    Using the estimate for the velocity of change of the concentration of
long living superheavy particles\cite{BBV}
$|\dot{n}_x| \sim 10^{-42}\, \mbox{cm}^{-3}\,\mbox{sec}^{-1} $,
and taking the life time $\tau_l $ of long living particles as
$2\cdot 10^{22} $\,sec, we obtain concentration
$n_X \approx 2\cdot 10^{-20} \,\mbox{cm}^{-3} $ at the modern epoch,
corresponding to the critical density for $M=10^{14} $\,Gev\,.

   Let us use the model with effective Hamiltonian~(\ref{HM})
where $\tau_l > t_U$ and take into account
transformations of the long living component into the short living
one due to the presence of baryon substance created by decays of
the short living particles in analogy with the regeneration
mechanizm for $K^0$-mesons.

    Let us investigate the model with the interaction which in the
basis   $\ |1 \rangle, \ |2 \rangle $  is described by the matrix
     \begin{equation}
H^d =     \left(
\begin{array}{cc}
0  & 0   \\
0  & - i \gamma \\
\end{array}        \right).
\label{Hd}
\end{equation}
    The eigenvalues of the Hamiltonian  $H+H^d$  are
     \begin{equation}
\lambda^d_{1,2} = E - \frac{i}{4}
\left(\tau_q^{-1} + \tau_l^{-1} \right) -i\,\frac{\gamma}{2} \pm
\sqrt{ \left( A - \frac{i}{4} \left(\tau_q^{-1} - \tau_l^{-1} \right)
\right)^2 -\frac{\gamma^2}{4} } \ .
\label{lamdop}
\end{equation}
    In case  when   $\gamma \ll \tau_q^{-1}$
for the long living component one obtains
     \begin{equation}
\lambda^d_{2} \approx  E - A - \frac{i}{2}\, \tau_l^{-1}
-i\,\frac{\gamma}{2} \,,
\label{ldolg}
\end{equation}
     \begin{equation}
\| \Psi_2(t) \|{}^2 = \| \Psi_2(t_0) \|{}^2 \exp \left[
\frac{t_0 - t}{\tau_l} - \int_{t_0}^t \gamma(t)\, d t \right].
\label{P21}
\end{equation}

    The parameter $\gamma$,  describing the interaction with the
substance of the baryon medium, is evidently dependent on its state
and concentration of particles in it.
   For approximate evaluations take this parameter as
proportional to the concentration of particles:
$\gamma = \alpha\, n^{(0)}(t)$.
       Putting
$\tau_l = 2 \cdot 10^{22}$\,sec, $t \le t_U$, $a(t)=a_0 \sqrt{t}$
by~(\ref{NbM})   one obtains
     \begin{equation}
\| \Psi_2(t) \|^2 = \| \Psi_2(t_0) \|^2 \exp \left[ \alpha 2 b^{(s)}
M^{3/2}\left( \frac{1}{\sqrt{t}} - \frac{1}{\sqrt{t_0}}
\right) \right].
\label{Ptt0}
\end{equation}
      So the decay of the long living component due to this mechanism
takes place close to the time  $t_0$.
    One can think that this interaction of $X_l$ with baryon charge
is effective for times, when the baryon charge becomes strictly
conserved, i.e. we take the time larger or equal to the electroweak
time scale, defined by the temperature of the products of decay
of $X_q$.
    This temperature is defined from
$ M n^{(s)}(\tau_q) \approx \sigma T^4 $
and is given by

     \begin{equation}
T(t) = \left( \frac{ 30\, b^{(s)} }{ \pi^2 N_l } \right)^{\!1/4}
\frac{ M^{5/8}\, \tau_q^{1/8} }{ k_B\, \sqrt{t} },
\label{T}
\end{equation}
   where $k_B$ is Boltzmann constant, $N_l$ is defined by the number of
boson $N_B$ and fermion $N_F$ degrees of freedom of all kinds of
light particles:
$N_l=N_B + \frac{7}{8} N_F$\, (see Ref.~\refcite{KKZ}).
   At time $t_X$ this temperature is equal to
     \begin{equation}
T(t_X) = \frac{64 \sqrt{\pi}}{3}
\left( \frac{ 30}{N_l} \right)^{\!1/4}\!\!
\left( b^{(s)} \right)^{\!5/4}\!
\left( M \tau_q \right)^{1/8}  \frac{ M^3}{k_B M_{Pl}^2}\,.
\label{TtX}
\end{equation}
   If $ \tau_q = 1/M $ and $N_l\sim 10^2$ -- $10^4 $, then for spinor
$X$-particles $ T(t_X) \approx 300$ -- $100 $\,Gev,
i.e. the electroweak scale for created particles
(which is however different from that for the background).

    So let us suppose $t_0 \approx t_X$.
    If  $d$ -- is the part of long living particles surviving up to
the time  $t$   $\ (t_U \ge t \gg t_C)$  then from~(\ref{d})
and (\ref{Ptt0})
one obtains the evaluation for the parameter~$\alpha$
     \begin{equation}
\alpha = \frac{ - 3 \ln d}{ 128 \pi (b^{(s)})^2} \,
\frac{M_{Pl}^2}{M^4} \,.
\label{ald}
\end{equation}
    For  $M=10^{14}$\,Gev    and  $d=10^{-15}$   one obtains
$\alpha \approx 10^{-30}$\,sm${}^3$/sec.
    If  $\tau_q \sim 10^{-38} - 10^{-35} $\,sec
then the condition  $\gamma(t) \ll \tau_q^{-1} $
used in Eq.~(\ref{ldolg}) is valid for $t> t_X$.
    For this value  $\alpha$ we have
$\gamma(t_U)\approx 10^{-36}$\,sec${}^{-1}$ $\ll \tau_l^{-1}$.
   So one can neglect this mechanism for the decay of the long living
component of $X$-particles for the modern epoch
while for early universe at $t_0 \approx t_X$ it was important.
    The entropy in this scenario is created due to decay of $X_l$
on quarks and antiquarks at the time $t_X$ when the Grand Unification
symmetry is totally broken.
    Baryon charge is created at $t_q $ which can be equal to
Compton time for $X$-particles $t_C \sim 10^{-38} - 10^{-35} $\, sec.

   Our scheme is the same for the scalar particles and the fermions.
   The superheavy fermions are used, for example, in some models of
neutrino mass generation (the {\it see-saw} mechanism) in
Grand Unification theories.\cite{GMRS,Yosh}
    New experiments on high energetic particles in cosmic rays
surely will give us more information on their structure and origin.


\end{document}